\documentclass[twocolumn,secnumarabic,amssymb, nobibnotes, aps, prd]{revtex4-2}

\usepackage{bm}
\usepackage{amsmath}
\usepackage{hyperref}
\usepackage{float}
\usepackage{graphics}
\usepackage{graphicx}
\usepackage{adjustbox}
\usepackage{subfigure}

\usepackage{adjustbox}
\usepackage{tikz}
\usepackage{color}
\usetikzlibrary{shapes}

\usepackage{grffile}
\usepackage{algorithm}
\usepackage{algpseudocode}
\algrenewcommand\algorithmicensure{\textbf{Output}}
\algrenewcommand\algorithmicrequire{\textbf{Input}}

\usepackage{hyperref}

\graphicspath{{./figures/}{./}}

\usepackage{cleveref}
\crefname{equation}{Eq.}{Eqs.}
\crefname{section}{Sec.}{Secs.}
\crefname{figure}{Fig.}{Figs.}

\makeatletter
\newcommand\footnoteref[1]{\protected@xdef\@thefnmark{\ref{#1}}\@footnotemark}
\makeatother

\newcommand{\be}{\begin{equation}}
\newcommand{\ee}{\end{equation}} 
\newcommand{\bea}{\begin{eqnarray}}
\newcommand{\eea}{\end{eqnarray}}

\newcommand{\f}[2]{\frac{#1}{#2}}

\newcommand{\ccup}[1]{\left\{#1\right\}}

\newcommand{\bup}[1]{\left(#1\right)}

\usepackage[normalem]{ulem} 

\newcommand{\mlot}{\mbox{{\small MultiOT}}}
\newcommand{\mlotsp}{\mbox{{\small MultiOTsp}}}
\newcommand{\spath}{\mbox{{\small SP}}}

\begin{document}

\title{Sustainable optimal transport in multilayer networks}%

\author{Abdullahi Adinoyi Ibrahim}%
\email[AI: ]{abdullahi.ibrahim@tuebingen.mpg.de}
\affiliation{Max Planck Institute for Intelligent Systems, Cyber Valley, T{\"u}bingen 72076, Germany}

\author{Daniela Leite}%
\email[DL: ]{daniela.leite@tuebingen.mpg.de}
\affiliation{Max Planck Institute for Intelligent Systems, Cyber Valley, T{\"u}bingen 72076, Germany}

\author{Caterina De Bacco}%
\email[CDB: ]{caterina.debacco@tuebingen.mpg.de}
\affiliation{Max Planck Institute for Intelligent Systems, Cyber Valley, T{\"u}bingen 72076, Germany}


\begin{abstract}
Traffic congestion is one of the major challenges faced by the transportation industry. While this problem carries a high economical and environmental cost, the need for an efficient design of optimal paths for passengers in multilayer network infrastructures is imperative. We consider an approach based on optimal transport theory to route passengers preferably along layers that are more carbon-efficient than the road, e.g., rails. By analyzing the impact of this choice on performance, we find that this approach reduces carbon emissions considerably, compared to shortest-path minimization. Similarly, we find that this approach distributes traffic more homogeneously, thus alleviating the risk of traffic congestions.
Our results shed light on the impact of distributing traffic flexibly across layers guided by optimal transport theory.
\end{abstract}

\maketitle

\section{Introduction}
Traffic congestion is a major problem in the transportation industry, with significant economic and environmental repercussions. The impacts of the environmental cost such as carbon emissions and other air pollutants, on public health can be sizable and need to be properly studied \cite{landrigan2018lancet}. Combining different transportation modalities, as in multilayer networks, can mitigate congestion and thus improve urban sustainability \cite{orozco2021multimodal}. 
Modeling traffic congestion on multilayer networks is crucial to investigate the efficiency and cost of operating such infrastructures \cite{wu2020traffic}.  Addressing this problem requires extracting what paths passengers take from source to destination, information that can then be used to analyze traffic patterns. Many route extraction methods are based on shortest-path minimization \cite{de2014navigability, lampo2021multiple, morris2012transport, sole2016congestion} or assignment strategy \cite{gao2019effective}. However, the shortest paths (i.e. selfish routing) might not always be the optimal path in a congested network \cite{lazar2020routing, po2021futility, mehr2018can}, hence the need for coordinated traffic congestion. In addition, empirical results have shown that passengers may not always consider the shortest route \cite{quercia2014shortest,zheng2018understanding}. While efforts have been made to go beyond shortest-path minimization using the cavity method or message-passing algorithms \cite{yeung2012competition,po2021futility,yeung2013physics,altarelli2015edge,de2014shortest}, these approaches are only valid in single-layer networks. In multilayer networks, several works focus more on analyzing the properties of passenger flows rather than proposing models to extract trajectories.  They consider  random walks \cite{de2014navigability} or shortest-path optimization \cite{morris2012transport,lampo2021multiple,sole2016congestion} to extract flows, thus necessarily influencing the results of subsequent analysis based on these strategies. 
Fewer models have been targeting transport optimization in multilayer networks \cite{wu2020traffic}. For instance, Ref.  \cite{gao2019effective} proposed a flow-assignment strategy on multilayer networks, while Ref. \cite{zhou2013efficient} developed a  recurrent algorithm for communication networks.\\

A principled and efficient approach for extracting optimal paths of passengers in networks is optimal transport (OT) theory \cite{bonifaci2012physarum,baptista2020network,baptista2020principled,lonardi2021designing}.
This approach has been applied recently to multilayer networks \cite{ibrahim2021optimal}, where the key idea is to flexibly tune between different cost functions in each of the different layers, thus capturing the specificity of each type of infrastructure. For instance, a road network is more sensitive to traffic congestion than a rail one, while the infrastructure of a rail network may be more costly to build. Our work builds from these ideas by adapting this model to study and evaluate optimal paths on multilayer networks under different scenarios. 
The goal of our work is to study the trajectories of optimal paths and compare them with those extracted from standard approaches relying on shortest-path minimization to identify key properties that are better optimized if one considers the multilayer character of the network. Our main contribution is threefold: 
First, we consider an optimal transport-based approach to extract optimal paths for passengers in multilayer networks, contrarily to standard approaches based on shortest-path minimization. 
Second, we propose a variant of this OT-based method that interpolates between OT and shortest-path minimization. While the extracted paths of the two OT-based models are longer than those obtained by shortest-path minimization, the rail layer is used by more passengers.
 Finally, we show that by using the optimal routes extracted by OT-based algorithms, passengers are more likely to encounter little or no traffic while less CO$_2$, leading to a reduced environmental cost. Our empirical results on synthetic and real data show the need for approaches that exploit the multilayer nature of multimodal transportation networks. 

\section{Optimal transport for traffic distribution in multilayer networks}\label{sec:model}
\begin{figure}[htbp]
	\centering
	\tikzset{every picture/.style={line width=0.75pt}} 
	
	\begin{tikzpicture}[x=0.3pt,y=0.3pt,yscale=-1,xscale=1]
		
		\draw  [color={rgb, 255:red, 189; green, 16; blue, 224 }  ,draw opacity=1 ] (219.47,610.59) -- (106.77,448.12) -- (104.98,127.22) -- (217.3,222.61) -- cycle ;
		\draw  [color={rgb, 255:red, 189; green, 16; blue, 224 }  ,draw opacity=1 ][line width=1.5]  (204.31,515.3) .. controls (199.66,515.31) and (195.89,511.53) .. (195.88,506.86) .. controls (195.88,502.2) and (199.64,498.41) .. (204.3,498.4) .. controls (208.95,498.4) and (212.72,502.18) .. (212.73,506.84) .. controls (212.73,511.51) and (208.97,515.3) .. (204.31,515.3) -- cycle ;
		\draw [color={rgb, 255:red, 189; green, 16; blue, 224 }  ,draw opacity=1 ]   (204.3,498.4) -- (194.54,373.67) ;
		\draw  [color={rgb, 255:red, 189; green, 16; blue, 224 }  ,draw opacity=1 ][line width=1.5]  (128.02,440.54) .. controls (123.37,440.54) and (119.6,436.76) .. (119.59,432.09) .. controls (119.59,427.43) and (123.35,423.64) .. (128.01,423.63) .. controls (132.66,423.63) and (136.43,427.41) .. (136.44,432.08) .. controls (136.44,436.74) and (132.68,440.53) .. (128.02,440.54) -- cycle ;
		\draw  [color={rgb, 255:red, 189; green, 16; blue, 224 }  ,draw opacity=1 ][line width=1.5]  (194.54,373.67) .. controls (189.89,373.67) and (186.11,369.89) .. (186.11,365.22) .. controls (186.1,360.56) and (189.87,356.77) .. (194.52,356.76) .. controls (199.17,356.76) and (202.95,360.54) .. (202.95,365.21) .. controls (202.96,369.87) and (199.19,373.66) .. (194.54,373.67) -- cycle ;
		\draw  [color={rgb, 255:red, 189; green, 16; blue, 224 }  ,draw opacity=1 ][line width=1.5]  (123.75,187.02) .. controls (119.1,187.02) and (115.32,183.25) .. (115.32,178.58) .. controls (115.31,173.91) and (119.08,170.12) .. (123.73,170.12) .. controls (128.38,170.11) and (132.16,173.89) .. (132.16,178.56) .. controls (132.17,183.23) and (128.4,187.01) .. (123.75,187.02) -- cycle ;
		\draw  [color={rgb, 255:red, 189; green, 16; blue, 224 }  ,draw opacity=1 ][line width=1.5]  (163.97,289.19) .. controls (159.31,289.2) and (155.54,285.42) .. (155.53,280.75) .. controls (155.53,276.08) and (159.3,272.29) .. (163.95,272.29) .. controls (168.6,272.28) and (172.38,276.06) .. (172.38,280.73) .. controls (172.38,285.4) and (168.62,289.19) .. (163.97,289.19) -- cycle ;
		\draw [color={rgb, 255:red, 189; green, 16; blue, 224 }  ,draw opacity=1 ]   (132.42,425.24) -- (188.52,371.26) ;
		\draw [color={rgb, 255:red, 189; green, 16; blue, 224 }  ,draw opacity=1 ]   (194.52,356.76) -- (168.38,289.19) ;
		\draw [color={rgb, 255:red, 189; green, 16; blue, 224 }  ,draw opacity=1 ]   (159.54,273.1) -- (128.16,184.6) ;
		\draw  [color={rgb, 255:red, 139; green, 87; blue, 42 }  ,draw opacity=1 ] (493.83,609.5) -- (381.13,447.03) -- (379.34,126.13) -- (491.66,221.52) -- cycle ;
		\draw  [color={rgb, 255:red, 139; green, 87; blue, 42 }  ,draw opacity=1 ] (402.38,439.44) .. controls (397.73,439.45) and (393.95,435.67) .. (393.95,431) .. controls (393.94,426.33) and (397.71,422.55) .. (402.36,422.54) .. controls (407.02,422.54) and (410.79,426.32) .. (410.8,430.98) .. controls (410.8,435.65) and (407.03,439.44) .. (402.38,439.44) -- cycle ;
		\draw  [color={rgb, 255:red, 139; green, 87; blue, 42 }  ,draw opacity=1 ] (468.89,372.57) .. controls (464.24,372.58) and (460.47,368.8) .. (460.46,364.13) .. controls (460.46,359.46) and (464.22,355.68) .. (468.88,355.67) .. controls (473.53,355.67) and (477.3,359.45) .. (477.31,364.11) .. controls (477.31,368.78) and (473.55,372.57) .. (468.89,372.57) -- cycle ;
		\draw  [color={rgb, 255:red, 139; green, 87; blue, 42 }  ,draw opacity=1 ] (398.1,185.93) .. controls (393.45,185.93) and (389.68,182.15) .. (389.67,177.49) .. controls (389.67,172.82) and (393.43,169.03) .. (398.09,169.03) .. controls (402.74,169.02) and (406.51,172.8) .. (406.52,177.47) .. controls (406.52,182.14) and (402.76,185.92) .. (398.1,185.93) -- cycle ;
		\draw  [color={rgb, 255:red, 139; green, 87; blue, 42 }  ,draw opacity=1 ] (438.32,288.1) .. controls (433.67,288.1) and (429.89,284.32) .. (429.89,279.66) .. controls (429.88,274.99) and (433.65,271.2) .. (438.3,271.2) .. controls (442.96,271.19) and (446.73,274.97) .. (446.74,279.64) .. controls (446.74,284.31) and (442.97,288.09) .. (438.32,288.1) -- cycle ;
		\draw [color={rgb, 255:red, 139; green, 87; blue, 42 }  ,draw opacity=1 ]   (406.78,424.15) -- (462.88,370.16) ;
		\draw [color={rgb, 255:red, 139; green, 87; blue, 42 }  ,draw opacity=1 ]   (468.88,355.67) -- (442.73,288.09) ;
		\draw [color={rgb, 255:red, 139; green, 87; blue, 42 }  ,draw opacity=1 ]   (433.89,272.01) -- (402.51,183.51) ;
		\draw  [color={rgb, 255:red, 139; green, 87; blue, 42 }  ,draw opacity=1 ] (478.67,514.21) .. controls (474.02,514.22) and (470.24,510.44) .. (470.24,505.77) .. controls (470.23,501.1) and (474,497.32) .. (478.65,497.31) .. controls (483.3,497.31) and (487.08,501.09) .. (487.08,505.75) .. controls (487.09,510.42) and (483.32,514.21) .. (478.67,514.21) -- cycle ;
		\draw [color={rgb, 255:red, 139; green, 87; blue, 42 }  ,draw opacity=1 ]   (478.65,497.31) -- (468.89,372.57) ;
		\draw  [color={rgb, 255:red, 189; green, 16; blue, 224 }  ,draw opacity=1 ][fill={rgb, 255:red, 255; green, 255; blue, 255 }  ,fill opacity=1 ] (201.86,469.43) .. controls (197.21,469.44) and (193.43,465.66) .. (193.43,460.99) .. controls (193.42,456.32) and (197.19,452.54) .. (201.84,452.53) .. controls (206.49,452.53) and (210.27,456.31) .. (210.27,460.97) .. controls (210.28,465.64) and (206.51,469.43) .. (201.86,469.43) -- cycle ;
		\draw  [color={rgb, 255:red, 189; green, 16; blue, 224 }  ,draw opacity=1 ][fill={rgb, 255:red, 255; green, 255; blue, 255 }  ,fill opacity=1 ] (198.6,419.54) .. controls (193.95,419.54) and (190.17,415.76) .. (190.17,411.09) .. controls (190.16,406.43) and (193.93,402.64) .. (198.58,402.63) .. controls (203.23,402.63) and (207.01,406.41) .. (207.01,411.08) .. controls (207.02,415.74) and (203.25,419.53) .. (198.6,419.54) -- cycle ;
		\draw  [color={rgb, 255:red, 189; green, 16; blue, 224 }  ,draw opacity=1 ][fill={rgb, 255:red, 255; green, 255; blue, 255 }  ,fill opacity=1 ] (160.48,406.7) .. controls (155.83,406.7) and (152.05,402.92) .. (152.05,398.26) .. controls (152.04,393.59) and (155.81,389.8) .. (160.46,389.8) .. controls (165.11,389.79) and (168.89,393.57) .. (168.89,398.24) .. controls (168.9,402.91) and (165.13,406.69) .. (160.48,406.7) -- cycle ;
		\draw  [color={rgb, 255:red, 189; green, 16; blue, 224 }  ,draw opacity=1 ][fill={rgb, 255:red, 255; green, 255; blue, 255 }  ,fill opacity=1 ] (181.46,331.43) .. controls (176.81,331.43) and (173.03,327.65) .. (173.03,322.98) .. controls (173.02,318.32) and (176.79,314.53) .. (181.44,314.52) .. controls (186.09,314.52) and (189.87,318.3) .. (189.87,322.97) .. controls (189.88,327.63) and (186.11,331.42) .. (181.46,331.43) -- cycle ;
		\draw  [color={rgb, 255:red, 189; green, 16; blue, 224 }  ,draw opacity=1 ][fill={rgb, 255:red, 255; green, 255; blue, 255 }  ,fill opacity=1 ] (148.48,250.17) .. controls (143.83,250.18) and (140.05,246.4) .. (140.05,241.73) .. controls (140.05,237.06) and (143.81,233.28) .. (148.46,233.27) .. controls (153.12,233.27) and (156.89,237.05) .. (156.9,241.71) .. controls (156.9,246.38) and (153.13,250.17) .. (148.48,250.17) -- cycle ;
		\draw  [color={rgb, 255:red, 189; green, 16; blue, 224 }  ,draw opacity=1 ][fill={rgb, 255:red, 255; green, 255; blue, 255 }  ,fill opacity=1 ] (137.22,218.8) .. controls (132.57,218.8) and (128.79,215.02) .. (128.79,210.35) .. controls (128.78,205.69) and (132.55,201.9) .. (137.2,201.9) .. controls (141.85,201.89) and (145.63,205.67) .. (145.63,210.34) .. controls (145.64,215) and (141.87,218.79) .. (137.22,218.8) -- cycle ;
		\draw  [color={rgb, 255:red, 189; green, 16; blue, 224 }  ,draw opacity=1 ][fill={rgb, 255:red, 255; green, 255; blue, 255 }  ,fill opacity=1 ] (130.96,366.89) .. controls (126.3,366.9) and (122.53,363.12) .. (122.52,358.45) .. controls (122.52,353.78) and (126.29,349.99) .. (130.94,349.99) .. controls (135.59,349.98) and (139.36,353.76) .. (139.37,358.43) .. controls (139.37,363.1) and (135.61,366.89) .. (130.96,366.89) -- cycle ;
		\draw  [color={rgb, 255:red, 189; green, 16; blue, 224 }  ,draw opacity=1 ][fill={rgb, 255:red, 255; green, 255; blue, 255 }  ,fill opacity=1 ] (145.35,328.24) .. controls (140.7,328.25) and (136.93,324.47) .. (136.92,319.8) .. controls (136.92,315.14) and (140.69,311.35) .. (145.34,311.34) .. controls (149.99,311.34) and (153.76,315.12) .. (153.77,319.78) .. controls (153.77,324.45) and (150.01,328.24) .. (145.35,328.24) -- cycle ;
		\draw  [color={rgb, 255:red, 189; green, 16; blue, 224 }  ,draw opacity=1 ][fill={rgb, 255:red, 255; green, 255; blue, 255 }  ,fill opacity=1 ] (187.8,257.38) .. controls (183.15,257.38) and (179.37,253.6) .. (179.37,248.93) .. controls (179.36,244.27) and (183.13,240.48) .. (187.78,240.47) .. controls (192.43,240.47) and (196.21,244.25) .. (196.21,248.92) .. controls (196.22,253.58) and (192.45,257.37) .. (187.8,257.38) -- cycle ;
		\draw  [color={rgb, 255:red, 189; green, 16; blue, 224 }  ,draw opacity=1 ][fill={rgb, 255:red, 255; green, 255; blue, 255 }  ,fill opacity=1 ] (158.16,459.52) .. controls (153.51,459.53) and (149.73,455.75) .. (149.73,451.08) .. controls (149.72,446.41) and (153.49,442.63) .. (158.14,442.62) .. controls (162.8,442.62) and (166.57,446.4) .. (166.58,451.06) .. controls (166.58,455.73) and (162.81,459.52) .. (158.16,459.52) -- cycle ;
		\draw [color={rgb, 255:red, 189; green, 16; blue, 224 }  ,draw opacity=1 ]   (196.63,455.76) -- (167.58,452.06) ;
		\draw [color={rgb, 255:red, 189; green, 16; blue, 224 }  ,draw opacity=1 ]   (160.14,442.62) -- (160.49,406.08) ;
		\draw [color={rgb, 255:red, 189; green, 16; blue, 224 }  ,draw opacity=1 ]   (155.05,393.42) -- (135.57,364.07) ;
		\draw [color={rgb, 255:red, 189; green, 16; blue, 224 }  ,draw opacity=1 ]   (134.95,350.79) -- (141.14,327.85) ;
		\draw [color={rgb, 255:red, 189; green, 16; blue, 224 }  ,draw opacity=1 ]   (145.34,311.34) -- (158.75,285.98) ;
		\draw [color={rgb, 255:red, 189; green, 16; blue, 224 }  ,draw opacity=1 ]   (153.77,319.78) -- (173.03,322.98) ;
		\draw [color={rgb, 255:red, 189; green, 16; blue, 224 }  ,draw opacity=1 ]   (156.9,241.71) -- (181.17,244.91) ;
		\draw [color={rgb, 255:red, 0; green, 0; blue, 0 }  ,draw opacity=1 ] [dash pattern={on 4.5pt off 4.5pt}]  (353.91,504.28) -- (470.24,504.16) ;
		\draw [color={rgb, 255:red, 0; green, 0; blue, 0 }  ,draw opacity=1 ] [dash pattern={on 4.5pt off 4.5pt}]  (202.95,365.21) -- (327.29,362.66) ;
		\draw [color={rgb, 255:red, 0; green, 0; blue, 0 }  ,draw opacity=1 ] [dash pattern={on 4.5pt off 4.5pt}]  (344.14,362.64) -- (460.46,364.13) ;
		\draw [color={rgb, 255:red, 0; green, 0; blue, 0 }  ,draw opacity=1 ] [dash pattern={on 4.5pt off 4.5pt}]  (132.16,178.56) -- (256.5,176.02) ;
		\draw [color={rgb, 255:red, 0; green, 0; blue, 0 }  ,draw opacity=1 ] [dash pattern={on 4.5pt off 4.5pt}]  (273.35,176) -- (389.67,177.49) ;
		\draw [color={rgb, 255:red, 0; green, 0; blue, 0 }  ,draw opacity=1 ] [dash pattern={on 4.5pt off 4.5pt}]  (313.57,278.17) -- (429.89,279.66) ;
		\draw [color={rgb, 255:red, 0; green, 0; blue, 0 }  ,draw opacity=1 ] [dash pattern={on 4.5pt off 4.5pt}]  (172.38,280.73) -- (296.72,278.19) ;
		\draw [color={rgb, 255:red, 0; green, 0; blue, 0 }  ,draw opacity=1 ] [dash pattern={on 4.5pt off 4.5pt}]  (136.44,432.08) -- (260.78,429.53) ;
		\draw [color={rgb, 255:red, 0; green, 0; blue, 0 }  ,draw opacity=1 ] [dash pattern={on 4.5pt off 4.5pt}]  (276.82,429.51) -- (393.95,431) ;
		\draw [color={rgb, 255:red, 0; green, 0; blue, 0 }  ,draw opacity=1 ] [dash pattern={on 4.5pt off 4.5pt}]  (212.73,506.84) -- (337.07,504.3) ;

		\draw (195.87,573.71) node [anchor=north west][inner sep=0.75pt]  [font=\large,rotate=-269.94]  {$\ \alpha $};
		\draw (465.9,570.66) node [anchor=north west][inner sep=0.75pt]  [font=\large,rotate=-269.94]  {$\ \gamma $};

	\end{tikzpicture}
	\caption{Multilayer structure with $N_{\alpha} = 15$ and $N_{\gamma} = 4$. The  network edges are represented by continuous lines (magenta and brown) and the two-edge path by dashed lines. The thicker magenta nodes represent stations belonging to both layers.} 
	\label{fig:multilayerstructure}
\end{figure}
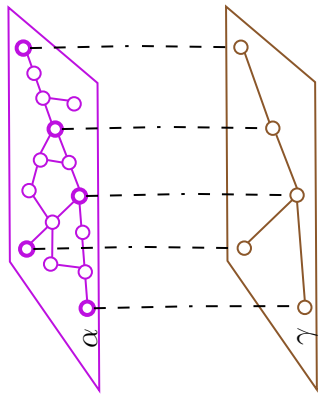

We denote a multilayer network as a graph denoted as $G(\{\mathcal{V}_{\alpha}\}_{\alpha}, \{\mathcal{E}_{\alpha}\}_{\alpha},\{\mathcal{E}_{\alpha}\gamma\}_{\alpha\gamma})$, where $\mathcal{V}_{\alpha}$, $\mathcal{E}_{\alpha}$, and $\mathcal{E}_{\alpha\gamma}$ denote the set of nodes, edges in layer $\alpha$, and inter-layer edges between layers $\alpha$ and $\gamma$; $\alpha=1,\dots, L$, where $L$ is the number of layers. We denote the number of nodes and edges as $N$ and $E$, and we assume that edges have length $l_e > 0$, which determines the cost of traveling through them. We consider the case of a two-layer network, but all results are valid for a higher number of layers. We denote the two layers as $\alpha, \gamma$ and consider a road network for $\alpha$ and a rail network for $\gamma$, as explained in detail in \Cref{sec:empirical}. We show an example of this structure in \Cref{fig:multilayerstructure}.

We consider passengers traveling through the networks and distinguish them by their origin and destination (traffic demands) stations $(o_{i},t_{i})$, where $o_{i},t_{i} \in \mathcal{V}= \cup_{\alpha}V_{\alpha}$. We denote as $\mathcal{S} = \ccup{(o_{i},t_{i})}$ the set of all origin-destination pairs, and $|\mathcal{S}|=M$ denotes their number.

We briefly describe the model of \cite{ibrahim2021optimal} to find optimal paths in multilayer networks using optimal transport theory.
It considers two main quantities on network edges: fluxes $F_{e}$ of passengers traveling through edge $e$ and conductivities $\mu_{e}$ determining $F_{e}$ passing through  an edge $e$. To keep track of the various routes that passengers have, a multi-commodity approach  is considered \cite{lonardi2021designing,lonardi2021multicommodity} in which passengers are distinguished based on their entry station $i\in \mathcal{S}$.
With this approach, the flux $F_{e}$ is an $M$-dimensional vector, where entries $F_{e}^{i}$ denote the flux of  passengers of type $i$ traveling on edge $e$. We assume the fluxes are determined by pressure potentials $p_{u}^{i}$ and $p_{v}^{i}$ defined on nodes as follows:
\be
	\label{eqn:flux}
	F_{e}^{i} := \frac{\mu_{e}}{l_{e}}\left({p_{u}^{i}-p_{v}^{i}}\right), \quad e=(u,v) \quad,
\ee
where $l_{e}$ is the length of edge $e$. Kirchhoff's law is imposed on network nodes to properly enforce mass conservation.
Finally, the dynamics assumes that the conductivity $\mu_{e}$ depends on flux $F_{e}$ as follows:
\begin{equation}
	\label{eqn:mu}
	\dot{\mu}_{e} = \mu_{e}^{\beta_{q_{e}}} \,\frac{\sum_{i \in \mathcal{S}}(p_{u}^{i}-p_{v}^{i})^{2}}{l_{e}^{2}} -\mu_{e}, \quad \forall e \in \mathcal{E}\quad,
\end{equation}
where $q_{e}$ encodes the layer to which the edge $e$ belongs. The parameter $0< \beta_{q_{e}} < 2$ determines the type of optimal transport problem one aims to solve: $0< \beta_{q_{e}} < 1$ discourage traffic congestion, $1< \beta_{q_{e}} < 2$ encourages path consolidation into few highways, while $\beta_{q_{e}} =1$ is shortest-path-like. Interpreting the conductivities as quantities proportional to the size of an edge, this dynamics enforces a feedback mechanism such that the edge size increases if the flux through that edge increases, and it decreases otherwise. 

It can be shown \cite{lonardi2021designing,ibrahim2021optimal} that the stationary solutions of \Cref{eqn:mu} minimize the multilayer transport cost function:
\be\label{eqn:J11}
J_{\beta} = \sum_{\alpha=1}^{L}\sum_{e\in \mathcal{E}_{\alpha}} l_{e}|| F_{e}||_{2}^{\Gamma(\beta_{\alpha})}\quad,
\ee
where $\Gamma(\beta_{\alpha}) = 2(2-\beta_{\alpha})/(3-\beta_{\alpha})$ for all $\alpha$, and the 2-norm is calculated over the $M$ entries of each $F_{e}$. Intuitively, solving the system of  \Cref{eqn:flux,eqn:mu} and Kirchhoff's law is equivalent to finding the optimal trajectories of passengers in a multilayer network, where optimality is given with respect to the transport cost in \Cref{eqn:J11}. We refer to this OT-based algorithm as \mlot \text{}. 

\subsection{\mlotsp: interpolating between OT and shortest-path minimization}
\begin{figure}
	\centering
	\includegraphics[width=1.0\linewidth]{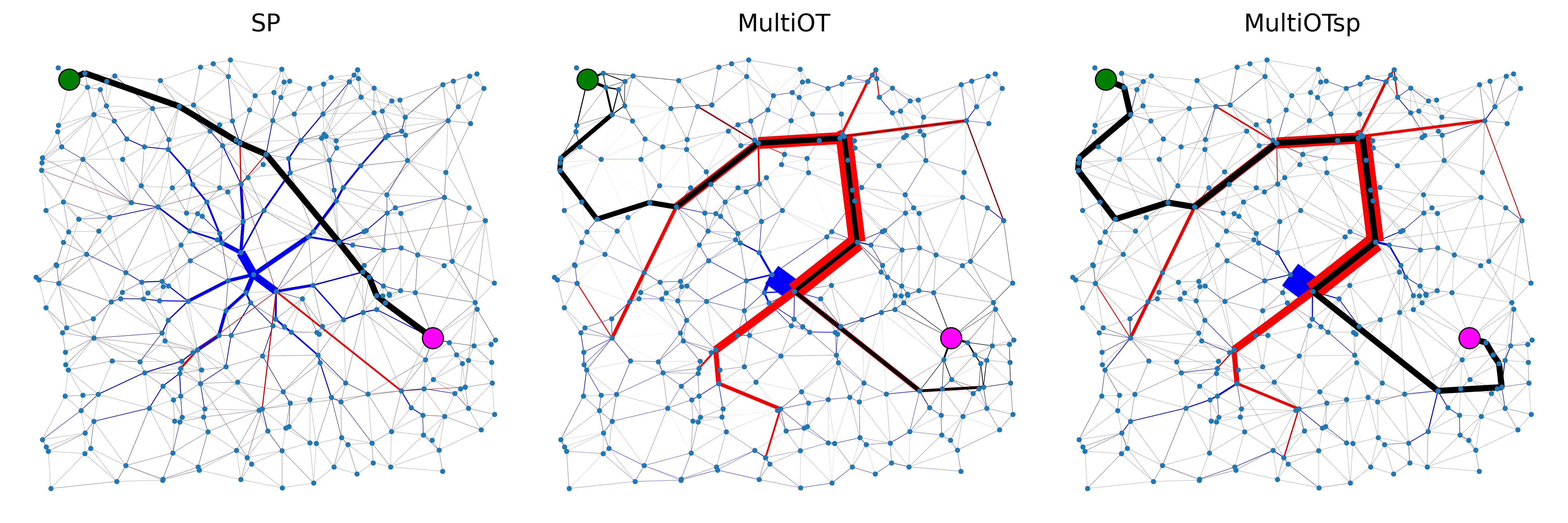}
	\caption{ Example trajectories. We show the trajectory of one type of passenger (black edges) whose origin and destination stations are the green and magenta nodes, respectively. We also highlight the total fluxes on edges, solutions of the OT problem including all other passengers, for a total of $M=300$. Blue and red edges denote road ($\alpha$) and rail ($\gamma$) layers, respectively.  Edge widths are proportional to the amount of passengers travelling through an edge. The exact width of the black edge has been either increased (for \spath{}) or reduced (for OT-based methods) in order to distinguish the flux of this type of passengers from the overall trajectories. Origin-destination pairs have been selected so that  $80\%$ of the passengers are directed towards a central node; $\beta_{\alpha}=0.5$ and $\beta_{\gamma}=1.9$. }
	\label{fig:allpaths}
\end{figure}

 The paths extracted by \mlot \text{} will encourage path consolidation along with the rail network and traffic minimization on the road network. Empirically, we observe that this model tends to distribute passengers of the same type (i.e. same origin and destination) along various routes, as shown in  \Cref{fig:allpaths}. While most of these passengers take the shortest among these routes, some distribute on longer ones to prevent traffic congestion.  This suggests an alternative algorithm that interpolates between \mlot \text{} and shortest-path minimization to select only the main relevant routes for each origin-destination pair among those extracted by \mlot. This can be done by inputting the solution of \mlot \text{} for each passenger type $i$ into a weighted shortest-path algorithm with edge weights defined as:
 \be
 w_{e}=\f{l_{e}}{|F_{e}^{i}|} \quad,
 \ee
  where the fluxes $F_{e}$ are those extracted from $\mlot$. All the passengers of type $i$ are then routed along the output path. We call this algorithm \mlotsp \text{} and show a pseudo-code in \Cref{alg:pipeline}. 
 The advantage of this weight function $w_{e}$ is that weakly used edges are consolidated on others that are on the optimal path (according to OT) of many passengers. These edges thus become more desirable when designing and individual ``consensus'' OT-based path that takes into account both path length \textit{and} optimal fluxes. The paths selected by \mlotsp \text{} rely strongly on how the fluxes are selected in the first place to determine the weights $w_{e}$. As the $F_{e}$ are calculated by considering all the passengers simultaneously (using \mlot),  the final optimal trajectories of \mlotsp \text{} are significantly distinct from those obtained by shortest-path minimization, which are independent from the surrounding environment.
We show an example of this in \Cref{fig:allpaths}.

\begin{algorithm}[H]
   \caption{\mlotsp}
   \label{alg:pipeline}
\begin{algorithmic}[1]
\Require{Graph $G(V,E,W)$, set $\mathcal{S}$ of origin-destination pairs, $\beta=(\beta_{1},\dots,\beta_{L})$}
\Ensure{Fluxes $\ccup{F_{e}}_{e}$}
    \Function{\mlotsp}{$G,\mathcal{S},\beta$}
 \State $\ccup{F_{e}}_{e} \gets $  \mlot($G,\mathcal{S},\beta$)
 \For{\texttt{$i = 1, \dots ,M$}}
    \State $\ccup{F_{e}^{i}}_{e} \gets $ weighted Dijkstra$(G,\mathcal{S},w)$ with $w_{e}= l_{e}/ |F_{e}^{i}|$ 
          \EndFor
           \State $F_{e} = (F_{e}^{1},\dots,F_{e}^{M})$, $\forall \,e$
   \EndFunction
  \end{algorithmic}
\end{algorithm}

In the following, we study the trajectories of optimal paths extracted by the three approaches: \mlot, \mlotsp \text{}, and shortest-path minimization (\spath). We use the implementation in \cite{github2021multiOT} for \mlot, while for \spath \text{} we use the Dijkstra algorithm \cite{dijkstra1959note}.

\section{Empirical results}\label{sec:empirical}
To investigate the relevant properties of the optimal paths extracted by the different algorithms, we simulate a variety of realistic traffic scenarios. 
	Specifically, we generate a dataset of synthetic 2-layer planar networks, where $\alpha$ simulates a road network and $\gamma$ simulates a rail network (e.g., a tram). 	The layer $\alpha$ is constructed first, by randomly placing $N_{\alpha}$ nodes in $[0,1] \times [0,1]$,
	and extracting its Delaunay triangulation \cite{guibas1985primitives}.
	We then select among them a subset of $N_{\gamma}$ nodes to build the layer $\gamma$ with an analogous procedure, thus ensuring that the two layers are connected. In total, in this construction the multilayer network has $N=N_{\alpha}$ nodes and resembles the situation in which all the stations in the second layer also have access to the road network. Notice that other constructions are possible, but this choice does not impact the validity of our model. 	
	 In our simulations, we set $N_{\alpha} = 300$ and $N_{\gamma} = 60$.
	We extract 20 different networks and $100$ random samples of origin-destination pairs for each of them, for a total of 2000 realizations for each parameters' configuration. With this, we aim at capturing different transportation scenarios in the two layers, as rail networks are less subject to traffic congestion but more costly to build, while we can state the opposite for road networks. \mlot \text{} (and thus \mlotsp) can capture these differences by suitably tuning $\beta$ in each layer: to discourage traffic congestion in the road layer, we set $\beta_{\alpha} = 0.5$ and vary $\beta_{\gamma}$ in $0 < \beta_{\gamma} < 2$ to study various scenarios.
	 In realistic scenarios, passengers have different origins and destinations; see \Cref{fig:allpaths} for an example. As we may expect in many urban scenarios that the most frequent destination is located in the city center, we assign to each passenger type its destination by default to be a central node. Then, to explore alternative scenarios where destinations are more heterogeneous, we consider a rewiring probability $p=[0,1]$ to rewire its destination at random. Specifically, for each passenger type, we rewire its destination to a random node with probability $p$. Hence, $p=0.0$ corresponds to having a monocentric destination where all passengers move towards a central node and $p=1.0$ corresponds to selecting all passengers' destinations at random. 
We consider $p = \{0.2,0.5,0.8\}$, but we show results for $p=0.5$, as the qualitative behavior is similar to that for the others, see \Cref{sec:apx}.
	These settings exhibit three important properties of the OT-based algorithms.


\section{Longer lengths but higher rail network usage}\label{sec:path length}
\begin{figure}
	\centering
	\includegraphics[width=0.9\linewidth]{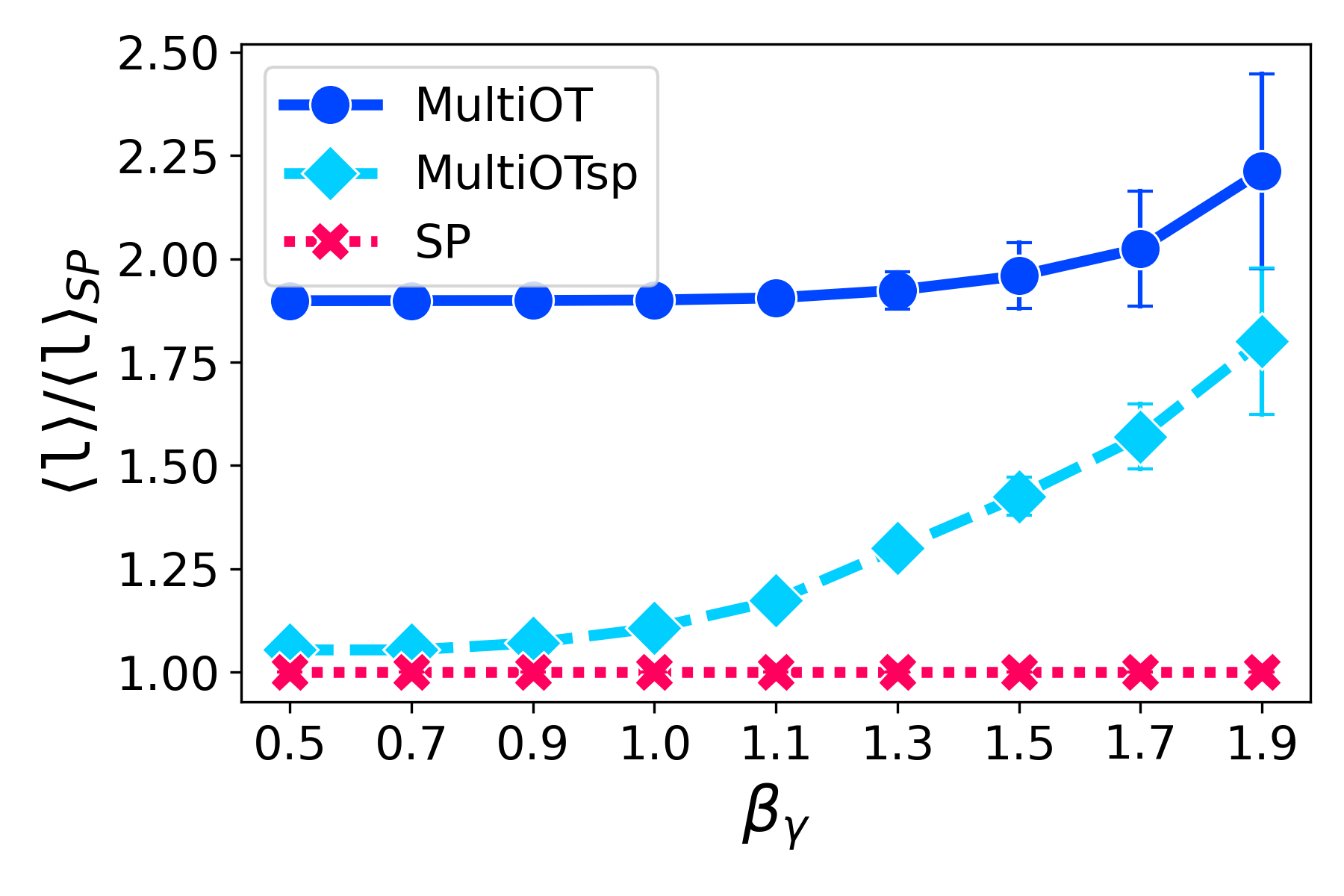}
	\caption{Average total path length ratio. We show the ratio of the average total path length to the one extracted from \spath. We set $p=0.5$, $\beta_{\alpha} = 0.5$ and vary $0 < \beta_{\gamma} < 2$. The results are averaged over 20 different network realizations with 100 randomly selected origin-destination pairs for each network realization. The markers and error bars are averages and standard deviations.}
	\label{fig:averagePL}
\end{figure}

Shortest-path optimization is utilized to minimize the total path length taken by passengers, hence we expect \mlot \text{} and \mlotsp \text{} to underperform \spath \text{} on this task. In fact, the performance of OT-based algorithms is expected to decrease as $\beta_{\gamma}$ increases, as shown in \Cref{fig:averagePL} by the average path length $\langle\, \l\,\rangle = \frac{1}{M} \sum_{e \in \mathcal{E} } l_{e} ||F_{e}||_{1}$ over the one obtained from a shortest-path algorithm.

\begin{figure}
	\centering
		\includegraphics[width=.95\columnwidth]{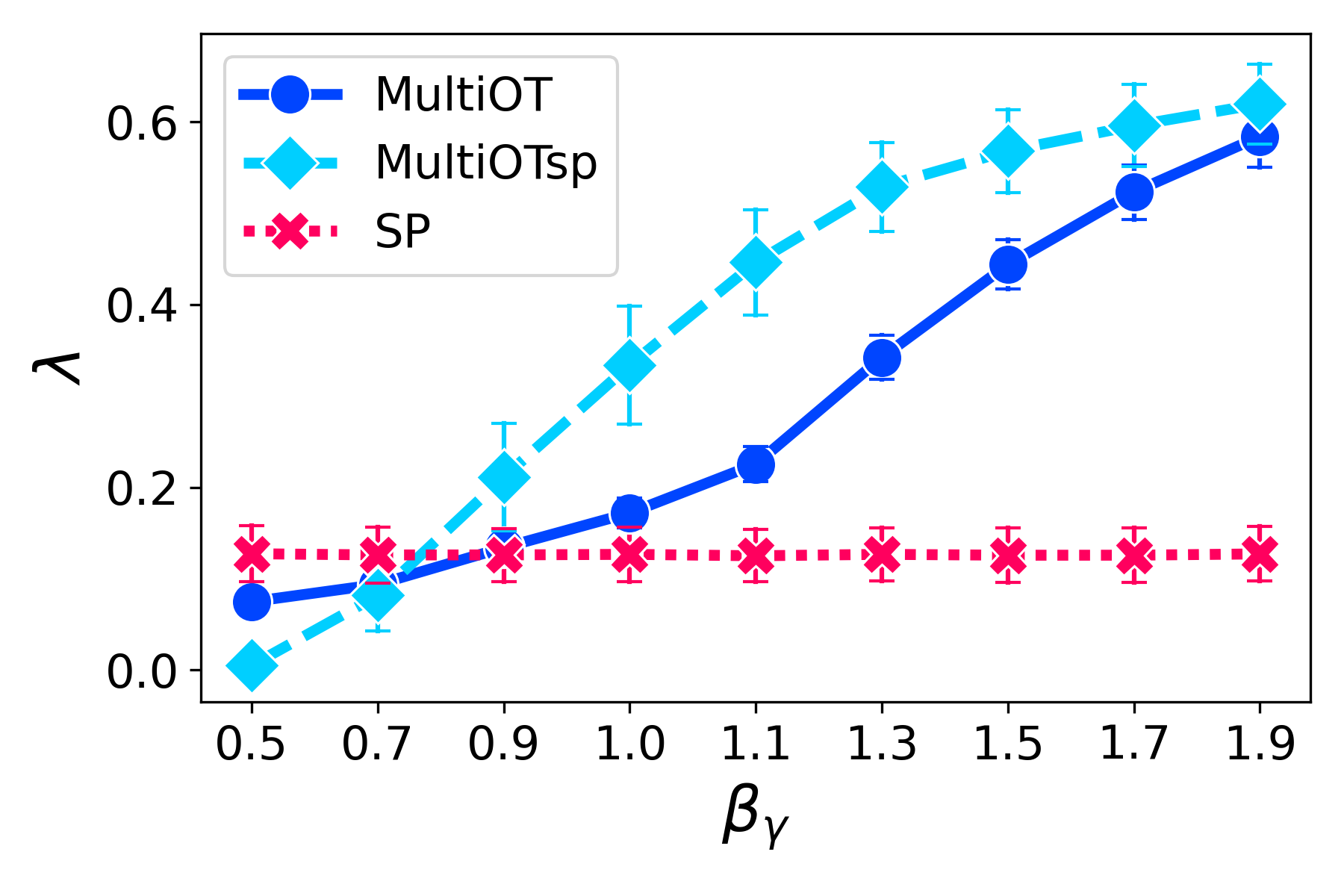}
	\caption{Coupling between layers. We show the coupling coefficient as defined in \Cref{eqn:couplingSyn}. All other settings remain the same as in \Cref{fig:averagePL}.}
	\label{fig:coupling}
\end{figure}

This is expected given that higher  $\beta_{\gamma}$ encourages more traffic to be routed towards the rail network at the cost of increased distance to cover, as the rail network has fewer and more distant nodes to reach than a road network. 
We then measure how passengers are distributed in the two layers by defining a coupling coefficient, a known concept to describe how well two layers are linked \cite{morris2012transport}.  We define:
\be\label{eqn:couplingSyn}
 \lambda =\f{1}{M}\sum_{i \in \mathcal{S}} \bup{\frac{\sum_{e \in \mathcal{E}_{\gamma}}|F_{e}^{i}|}{\sum_{e \in \mathcal{E}_{\alpha} \cup \mathcal{E}_{\gamma}} |F_{e}^{i}| }} \quad,
\ee
where the numerator inside the parentheses contains only the flux in the rail layer so we can distinguish how many passenger types effectively use that layer in their trajectories. This definition is valid for two-layer networks, such as the empirical networks studied here. However, one can appropriately generalize it for networks with more than two layers.
The usage of the rail layer increases monotonically for both OT-based algorithms, as shown in \Cref{fig:coupling}, with \mlotsp \text{} reaching higher usage values. This suggests that the shortest-path routes selected from the possible paths output by \mlot \text{} are composed of a significant amount of rail edges. This also shows that the raw solutions output of \mlot \text{} consider paths more distributed across the road layer, as qualitatively observed in \Cref{fig:allpaths}.

\section{Multilayer optimization can decrease carbon consumption}\label{sec:carbon usage}

\begin{figure}
	\centering
	\includegraphics[width=0.95\linewidth]{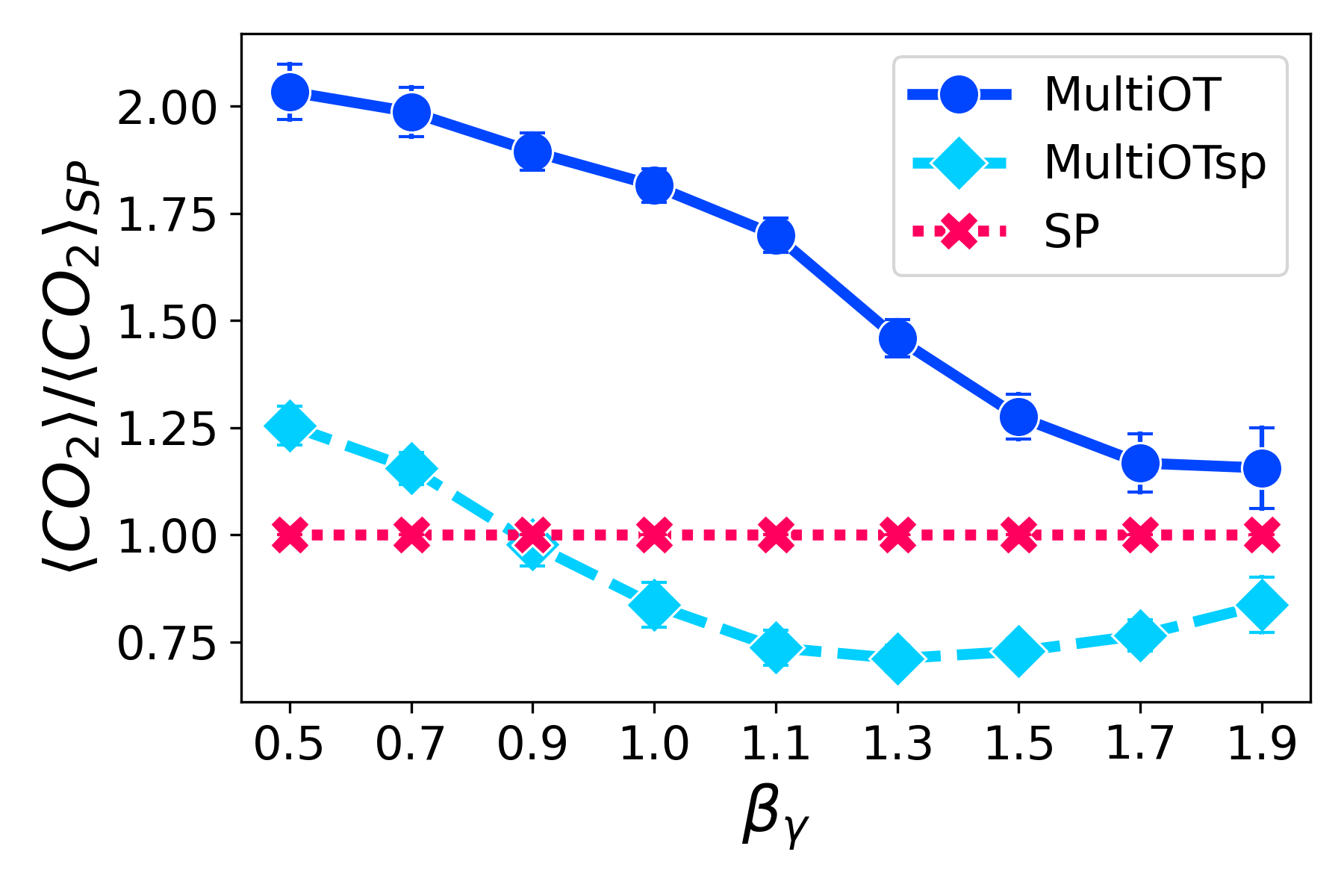}
	\caption{Carbon emission ratio. We show the ratio of the average carbon emissions as defined in \Cref{eqn:COave} to the one obtained by \spath. All other settings remain the same as in \Cref{fig:averagePL}.}
	\label{fig:emission}
\end{figure}

As more passengers take longer paths while being encouraged to use the rail network, they also consume less carbon per unit of length. The question is whether the increased length can be properly compensated by the decrease in carbon consumption. We tested this on the same paths extracted to plot \Cref{fig:averagePL} by measuring the average $CO_{2}$ consumption per passenger as:
\be\label{eqn:COave}
\langle\, CO_{2}\,\rangle = \frac{1}{M} \sum_{e \in \mathcal{E} } r_{q_{e}} l_{e} ||F_{e}||_{1}\quad,
\ee
where $r_{\alpha}$ is the carbon emission rate in layer $\alpha$. This has a dimension of unit of mass (e.g. $g$) per passenger per unit of length (e.g. pkm). For instance, a bus on average generates $101.87g/pkm$ \cite{defra2017department} while a train generates $28.39g/pkm$ \cite{eea2017carbon}. Hence, defining $r_{\alpha}$ as the rate of the road layer and considering buses traveling on it, we can set $r_{\gamma}= 28.39r_{\alpha}/101.87 = 0.28 \, r_{\alpha}$. These values can be changed accordingly with more specific values if a traffic manager has precise statistics of vehicles' types traveling on the network. By leveraging optimal transport with a bias towards shortest paths, \mlotsp \text{} is able to decrease carbon consumption the most compared to \spath, measured by the ratio of its $\langle\, CO_{2}\,\rangle$ over that produced by \spath. A minimum is reached for $1.1\leq \beta_{\gamma}\leq 1.5$ where \mlotsp \text{} produces $25\%$ fewer emissions than a shortest-path routing algorithm, as shown in \Cref{fig:emission}. 
This important result is a consequence of flexibly tuning the cost to be optimized in each layer, as allowed by $\beta$ in  \Cref{eqn:J11}.  In particular, $\beta_{\gamma}>1$ encourages paths to consolidate on the rail layer, while $\beta_\alpha=0.5$ controls for traffic congestion on the road layer. The fact that the minimum consumption of \mlotsp \text{} has not been realized at the highest value $\beta_{\gamma}=1.9$, where the paths are consolidated into the fewest rail routes,  further suggests that there is a trade-off between keeping the path lengths short while directing more passengers towards the rail layer. In fact, at $\beta_{\gamma}=1.9$, as the number of passengers redirected towards the second layer increases, they also have to take longer routes, thus emitting more carbon. A value of $\beta = 1.3$ results in a nice tradeoff between these two competing behaviors in terms of carbon emission.
On the contrary, \mlot \text{} shows a monotonic decreasing behavior with a minimum reached at $\beta=1.9$, but still higher than that emitted by \spath. This is a consequence of the higher number of possible paths that passengers can take as routed by \mlot, which are by default longer than those obtained by \mlotsp{} and use more edges of the road layer. As a consequence, the longer length does not seem to justify the higher usage of the rail layer.
 
\section{Traffic congestion}\label{sec:carbon emission}
\begin{figure}
	\centering
	\includegraphics[width=.95\columnwidth]{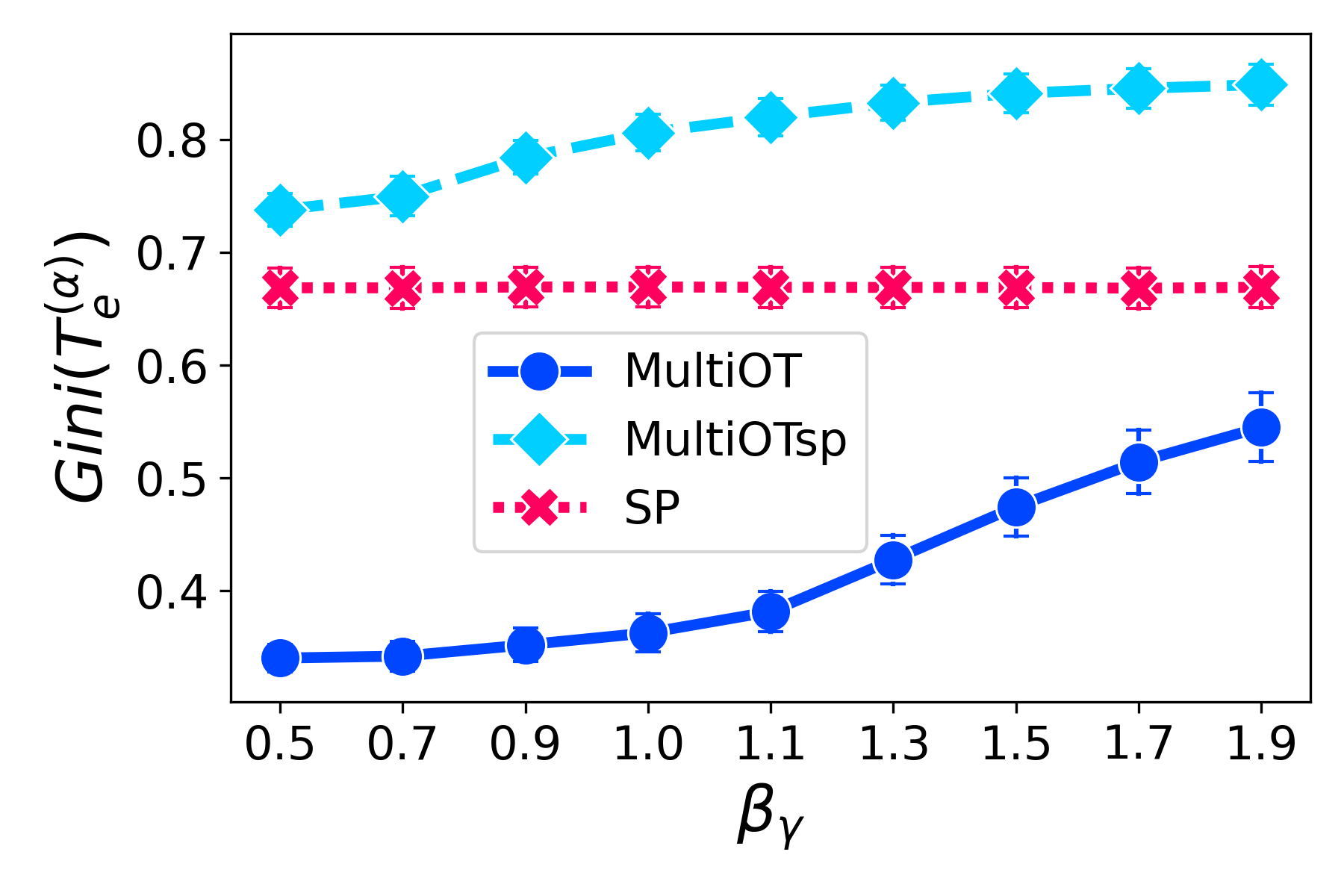}
			\includegraphics[width=1\columnwidth]{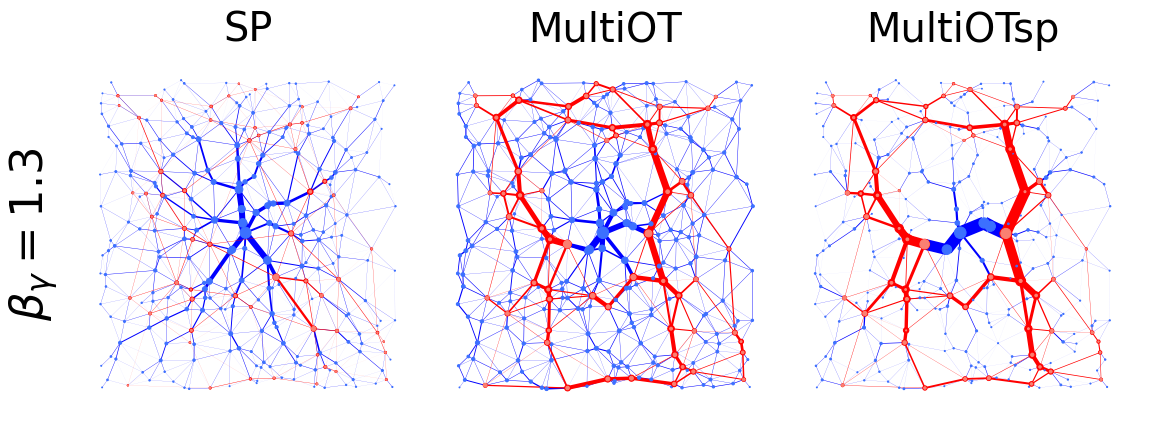}\vspace{-0.35cm}
	\includegraphics[width=1\columnwidth]{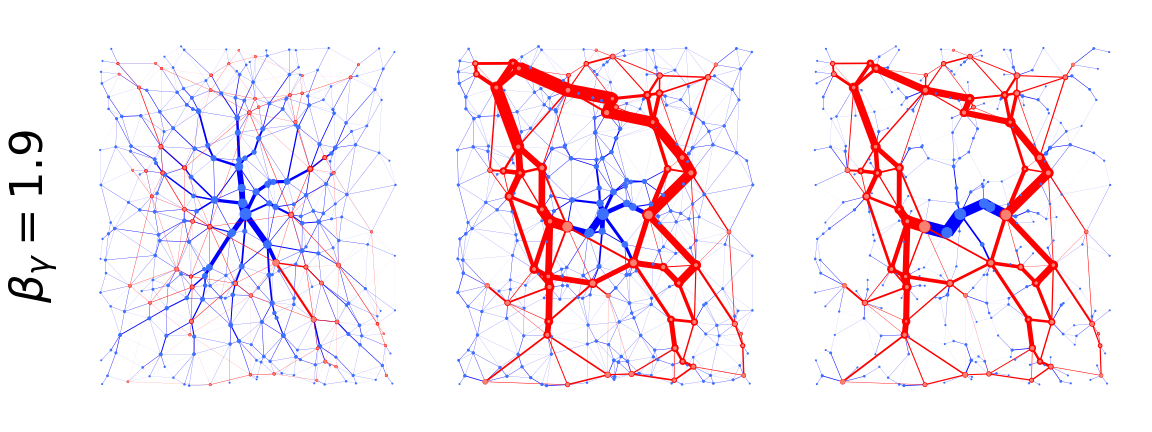}
	\caption{Traffic distribution. Top) Gini coefficient of the traffic on the road layer $\alpha$. Bottom) optimal trajectories. Edge widths are proportional to $||F_{e}||_{1}$ averaged over 10 samples of origin-destination configurations; $p=0.5$, $\beta_{\alpha}=0.5$. When $\beta_{\gamma}>1$, OT-based methods are consolidated into fewer edges in layer $\gamma$. The \spath{} on the other hand is not affected by this parameter, showing flows of passengers on more edges compared to OT-based which consolidate into fewer edges. Blue and red edges correspond to road ($\alpha$) and rail ($\gamma$) layers, respectively.   }
	\label{fig:traffic}
\end{figure}

All the results of the previous section were interpreted with the assumption that the flow of passengers is regular, even on high-traffic edges. Instead, if we account for traffic slowing down the flow on edges with a high density of travelers, those vehicles emit more carbon while they keep their engines on longer. 
  The routes suggested by \mlot \text{} are less sensitive to this, hence we also expect a lower carbon emission than that shown in \Cref{fig:emission} when accounting for traffic. We thus measure traffic load on edges as
  \begin{eqnarray}\label{eqn:Tedge}
T_{e} = \frac{1}{ n }\vert \vert F_{e} \vert \vert_{1} \quad,
\end{eqnarray}
where $n$ is the total number of passengers and measures the Gini coefficient Gini$(T_{e})\in [0,1]$ as a global network metric of inequality of how traffic is distributed on the network \cite{dixon1987bootstrapping}, with a Gini close to 1 meaning high inequality in flow assignment along edges. As the road layer is the one more sensitive to potential traffic bottlenecks, we consider only the traffic on road edges and denote with Gini$(T_{e}^{(\alpha)})$ the Gini coefficient calculated using only $e \in \mathcal{E}_{\alpha}$. As expected, \mlot \text{} has more balanced traffic than the other two algorithms, as shown in \Cref{fig:traffic}. While congestion increases with $\beta_{\gamma}$, even at the maximum $\beta=1.9$ the Gini coefficient is lower than that of \spath. The reason for this increase is that paths consolidate more on those fewer road edges that allow a connection to the rail layer, as can be seen on the example optimal routes in \Cref{fig:traffic}, a behaviour also observed in previous studies \cite{chodrow2016demand,strano2015multiplex}.
This is exacerbated in \mlotsp, as one can notice that central road edges are overly trafficked when many passengers exit the rail to reach the final destination in the center. This also causes the Gini coefficient of \mlotsp \text{} to be higher than that of \spath. In other words, few central edges cause most of the traffic for \mlotsp.  This can be partially alleviated by increasing $p$ towards 1 as fewer destinations are directed towards the network center, although this may become an unrealistic assumption in urban scenarios. Alternatively, one can simply add rail stations in the center, so that passengers do not have to commute one extra mile to reach their final destination, a scenario that we explore below in the case of a real network.
 
\section{Real multilayer network}\label{sec:real network}
Next, we examine these properties on a real 2-layer network of the city of Bordeaux \cite{kujala2018collection}, where  $\alpha$ and $\gamma$ represent the bus and tram networks, respectively. Similar to the synthetic network, we compare the performances of OT-based algorithms with SP on this network.  We set $p  = 0.2$ to consider the situation where the majority of passengers are directed towards the city center, a central node coinciding with a tram station,  and we extract 100 realizations of origin-destination pairs. 

The tram in this city travels through its own reserved lanes, independently from other vehicles. Hence, although the two layers are physically located next to each other, there is no mixing of fluxes from the two layers on edges.  This may differ in other real situations. While in principle our model is best suited for independent usage of the space by the various layers (e.g. road and subway or the case studied here), the results shown here may still apply if we assume that the physical presence of the tram only marginally impacts the traffic in the road layer, compared to other vehicles. Specifically, the combination of high enough capacity (number of passengers that can fit into a tram) and lower frequency of trams than other vehicles on the road may allow us to assume independence between the tram and road layer. In fact, while the tram may have many passengers traveling at any given time along an edge, these are all entering inside the same wagons. Thus, the space occupied by the tram is limited by its physical shape.  Instead, the same amount of passengers would need to distribute in many different cars, thus occupying much more space, potentially creating congestion. 
In general, in situations in which trams have reserved lanes that cars cannot enter, our treatment applies without any further assumption. 

\begin{figure}
	\centering
	\includegraphics[width=0.95\linewidth]{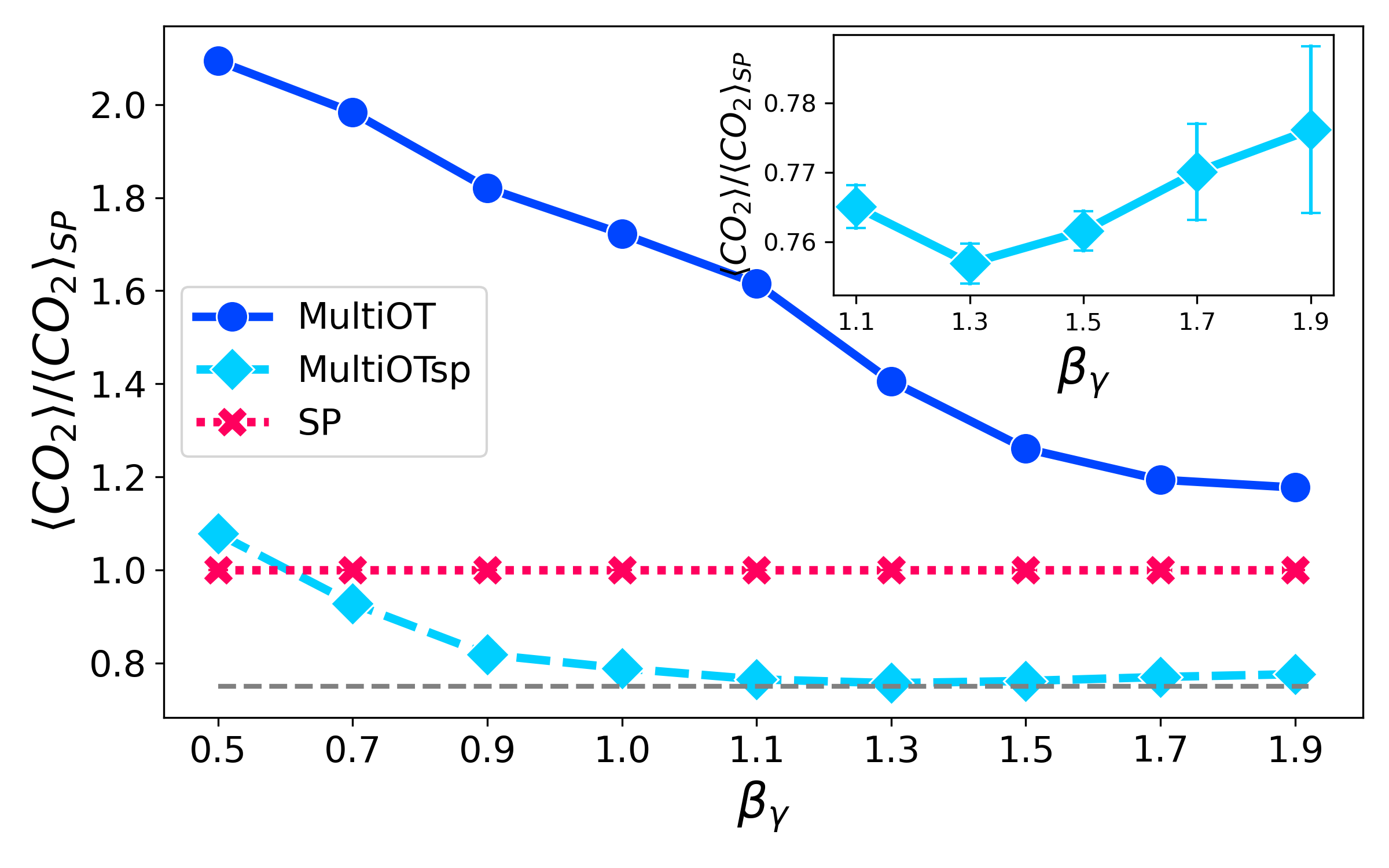}
	\caption{Carbon emissions on the Bordeaux network. Ratio of the average carbon emission over that of \spath. Here, we set $p=0.2$, thus favoring monocentric destinations. The grey-dashed line shows the minimum value obtained by \mlotsp{}, corresponding to $\beta_{\gamma}=1.3$.
	Inset:  zoom-in of \mlotsp{} values for $1.1 \le \beta_{\gamma}\le 1.9$. }
	\label{fig:emission_realnet}
\end{figure}

\begin{figure*}[htbp]
	\centering
	\includegraphics[width=.29\textwidth]{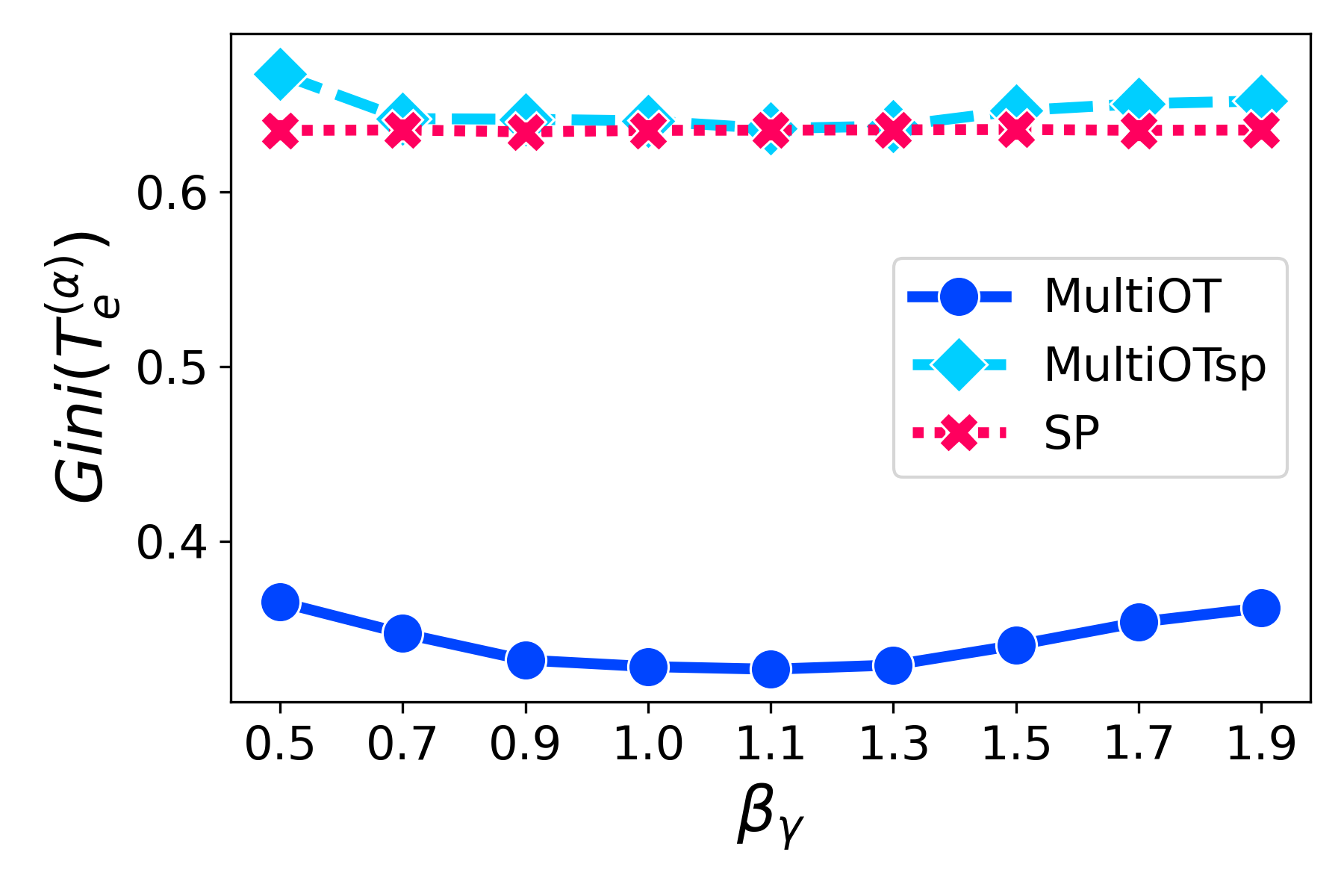}
	\includegraphics[width=.70\textwidth]{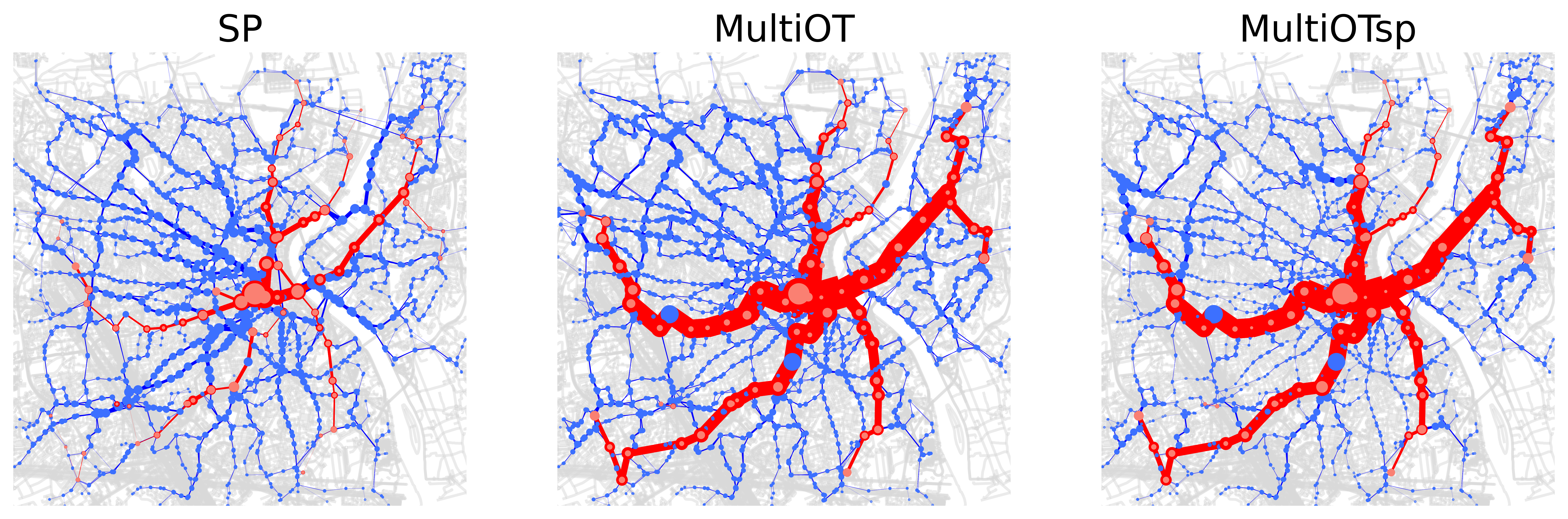}
	\caption{ Traffic distribution on the Bordeaux network. We set $p=0.2$ and $\beta_{\alpha} = 0.5$. Left) Gini coefficient calculated using the traffic on the road network. Error bar (standard deviation) are smaller than marker size. Center-Right) Traffic distribution for each of the algorithms with $\beta_{\gamma}=1.9$. Red and blue edges denote tram and road layers, respectively. Node and edge sizes are proportional to the amount of passengers traveling through them.  The results are averaged over 100  samples of origin-destination pairs.}
	\label{fig:traffic_realnet}
\end{figure*}

We find that \mlotsp \text{} produces $25\%$ fewer carbon emissions than \spath \text{} for $1.1 \le \beta_{\gamma} \le 1.5$, as shown in \Cref{fig:emission_realnet}, similar to what observed on synthetic networks. \mlot \text{} has a minimum at $\beta_{\gamma}=1.9$, but the emissions are higher than \spath. We argue that also in this case this is due to the assumption that the flow of vehicles is smoothly moving, with no traffic congestion causing velocity to decrease, and thus causing emissions to increase nearby traffic bottlenecks. 
To assess this hypothesis, we investigate the distribution of fluxes on the road layer by measuring the traffic $T_{e}$ on edges in layer $\alpha$. We find that indeed \mlot \text{} has path trajectories more homogeneously distributed across the road layer as measured by the Gini coefficient plotted in \Cref{fig:traffic_realnet} along with an example solution, potentially lowering the number of traffic jams. As we can see from an example solution in the same figure, the two OT-based variants distribute passengers in higher amounts along the tram network, thus lowering the road's usage, while \spath \text{} makes use of the tram mainly in the vicinity of the central node. We can further notice how \mlot \text{} uses the road with higher intensity than \mlotsp, but the road edges have less traffic than those obtained by  \spath. As for \mlotsp, the road edges with the most traffic are those nearby tram stations, like those in the upper left corner in the figure.

To better quantify the potential impact of traffic congestion as a proxy for the potential increase in CO$_{2}$ emission, we consider a measure of transport cost used before in similar problems \cite{yeung2012competition,yeung2013physics} and defined as
	\be\label{eqn:Jalpha}
	J_{\alpha} = \sum_{e\in \mathcal{E_{\alpha}}} l_{e}  \vert \vert F_{e} \vert \vert_{1}^{2} \quad, 
	\ee 
where the exponent $2$ discourage traffic. This should not be confused with the definition of \Cref{eqn:J11}, which is the one used to extract the optimal paths in our model, i.e. to solve the OT problem. Specifically, \Cref{eqn:J11}  uses different $\beta$ for edges in different layers. In particular, it allows us to encourage both path consolidation in one layer and path distribution in another. Instead, \Cref{eqn:Jalpha} only discourage traffic, as the exponent is greater than 1. In addition,  \Cref{eqn:J11} considers the norm-2 of the passengers' flows, while \Cref{eqn:Jalpha}  considers norm-1. While the latter is more intuitive, as it is the total number of passengers traveling along an edge, the former admits rigorous theoretical guarantees for OT to converge to an optimal solution. This does not apply to a cost function using norm-1; see  \cite{lonardi2021multicommodity} for a detailed discussion. 

\begin{figure}
	\centering
	\includegraphics[width=0.95\linewidth]{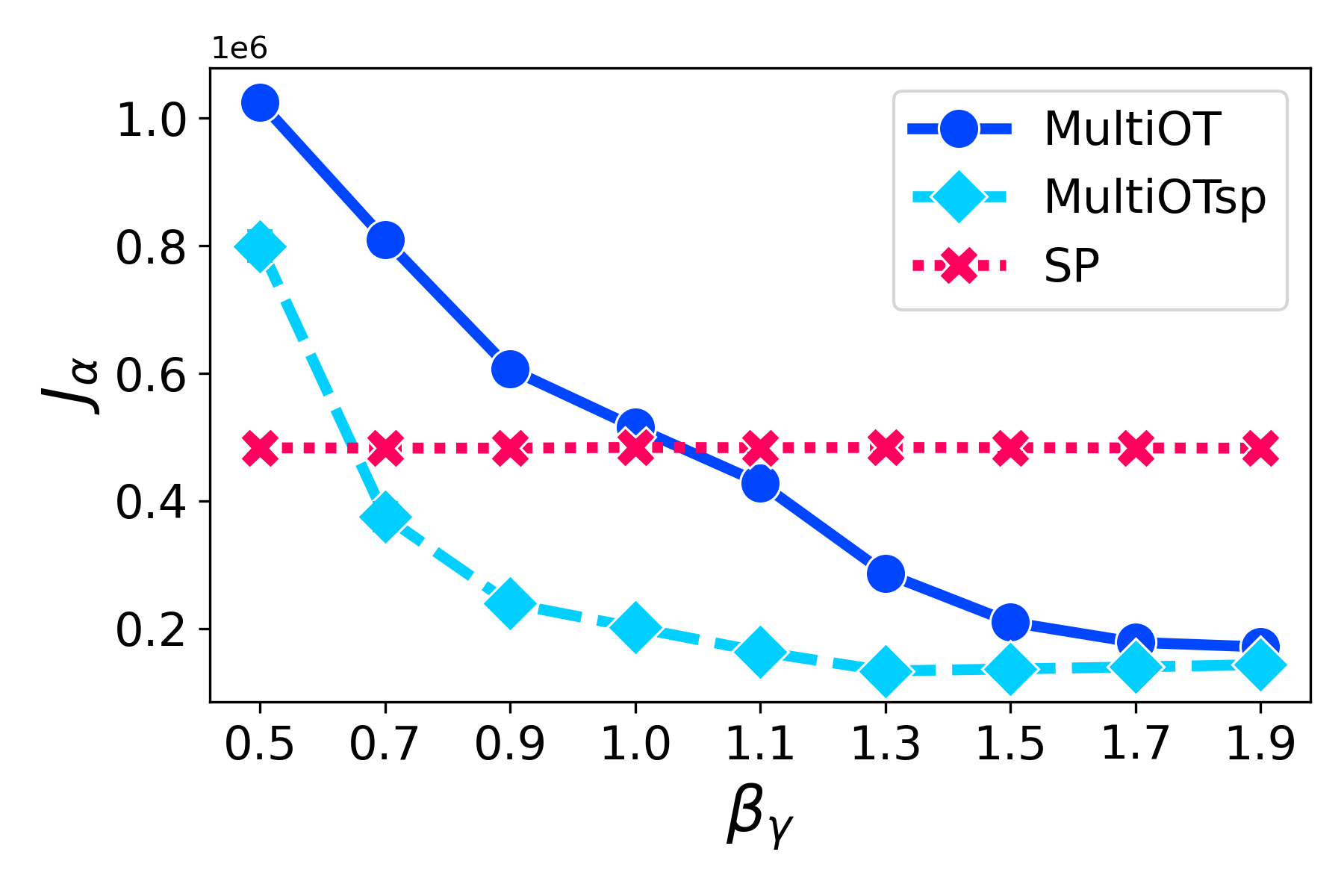}
	\caption{Transport cost on the road layer $\alpha$ of Bordeaux network. The cost is defined as in \Cref{eqn:Jalpha}; here $p=0.2$.}
	\label{fig:jl1bordeauxbustram0}
\end{figure}

As seen in \Cref{fig:jl1bordeauxbustram0}, both OT-based algorithms outperform  SP as $\beta_{\gamma}$ increases, meaning that passengers traveling on paths generated by the OT-based algorithms will generally record less road traffic congestions compared with the paths extracted by \spath. Assuming that velocity decreases along congested edges, we conjecture that this would result in \mlot \text{} having lower carbon emissions than \spath. 

\section{Discussion \& Conclusion}\label{sec:conclusion} 

Designing and extracting optimal passenger flows in a transportation network is crucial for reducing traffic congestion and environmental costs. Methods based on shortest-path optimization are optimal in terms of reducing the average shortest path length to reach destination, but they may fail in terms of other relevant transportation metrics. In addition, passengers do not always follow the shortest route \cite{quercia2014shortest}, hence the need for alternative approaches to extract path trajectories and investigate their properties in multilayer networks. 
 We present two models based on optimal transport theory that can flexibly tune the amount of traffic routed in the different layers to encourage usage of rail networks while reducing traffic on the road. As a result, optimal trajectories extracted with these methods significantly decrease the amount of carbon emissions compared to shortest-path minimization, while also being more robust to traffic congestions.
In particular, we found that \mlotsp \text{}, by interpolating between optimal transport and shortest-path minimization, can achieve the lowest amount of carbon emissions under the hypothesis of smooth flow of passengers in a network. Instead, \mlot \text{}, based purely on optimal transport, distributes paths more homogeneously, thus being potentially more robust against increased carbon emissions when accounting for passengers' flow slowing down along traffic bottlenecks. This can be tested quantitatively in real scenarios by having access to empirical data of different velocities during traffic congestion, along with detailed velocity limits imposed by regulations in different parts of the network. One could potentially compare the theoretical results with the empirical ones observed from real data as in \cite{taillanter2021empirical}.\\
In general, we show that models based on optimal transport can be used to design optimal routes for passengers in a multilayer network, and we investigate scenarios beyond those obtained by using standard shortest-path algorithms.
In this work, we assumed fixed origin-destination pairs, but one can further generalize this analysis by considering dynamical traffic demands that change in time. This would require suitably adapting the models studied in this work to account for this, for instance borrowing ideas from  \cite{yeung2019coordinating,lonardi2021infrastructure,hu2013adaptation, corson2010fluctuations, hu2012biological}. Similarly, we did not explore here the possibility of traffic diversions due to road blockages or changing conditions in the network structure \cite{tai2021optimally,kleineberg2017geometric,wandelt2018comparative,wang2018group}. Studying the robustness of the methods investigated in this work to these scenarios would be an interesting subject for future work. Finally, it would be interesting to investigate more complex scenarios with more than two layers, possibly on a larger scale than that of a unique urban scenario. To facilitate future analysis, we provide an open source implementation of our code at \cite{github2021multiOT}.

\subsection*{Acknowledgments}
The authors thank the International Max Planck Research School for Intelligent Systems (IMPRS-IS) for supporting A.A.I and D.L.


\appendix

\section{Additional results varying $p$}\label{sec:apx}
\begin{figure}[htbp]
	\centering
	\subfigure[]{\includegraphics[width=.47\linewidth]{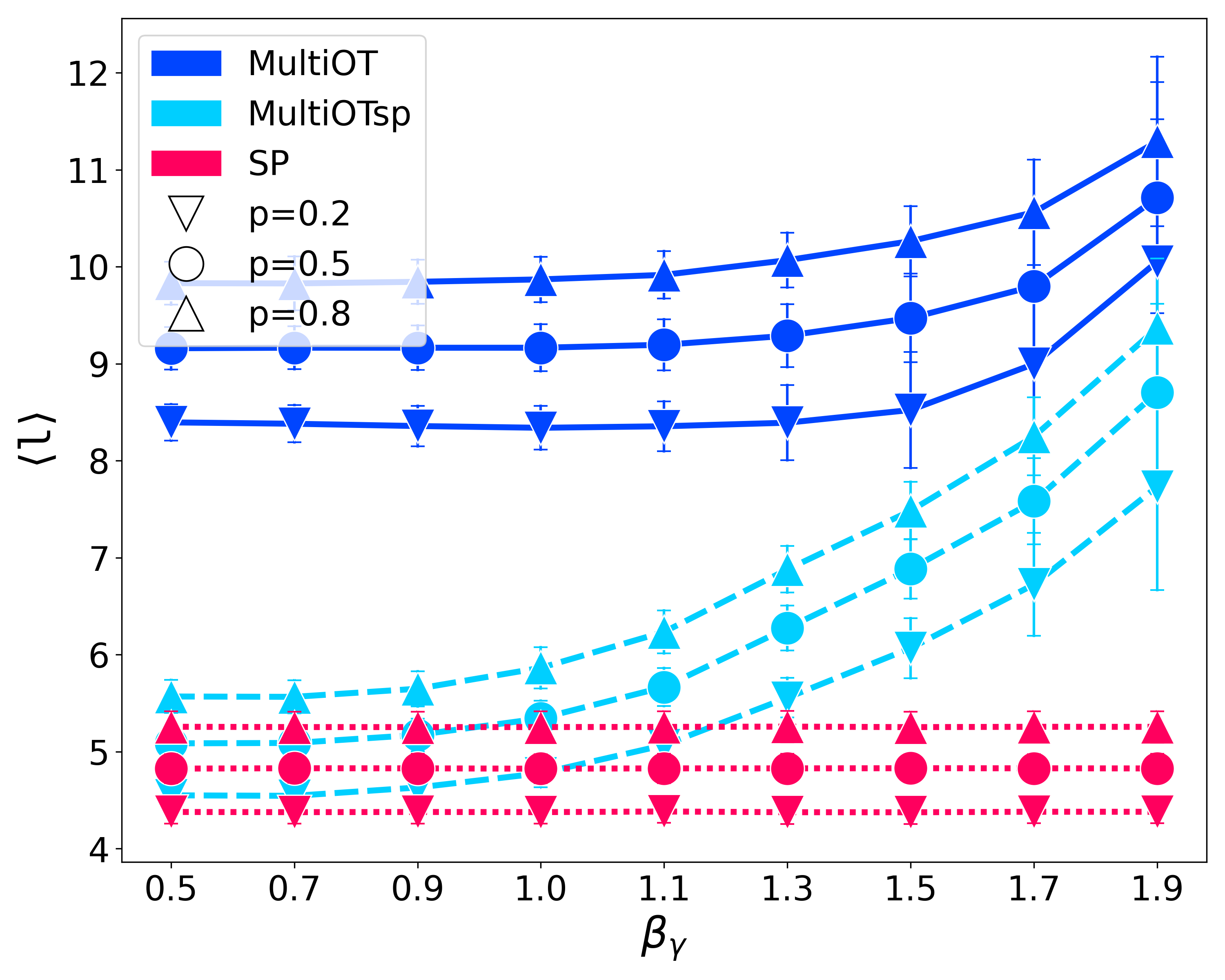}}
	\subfigure[]{\includegraphics[width=.47\linewidth]{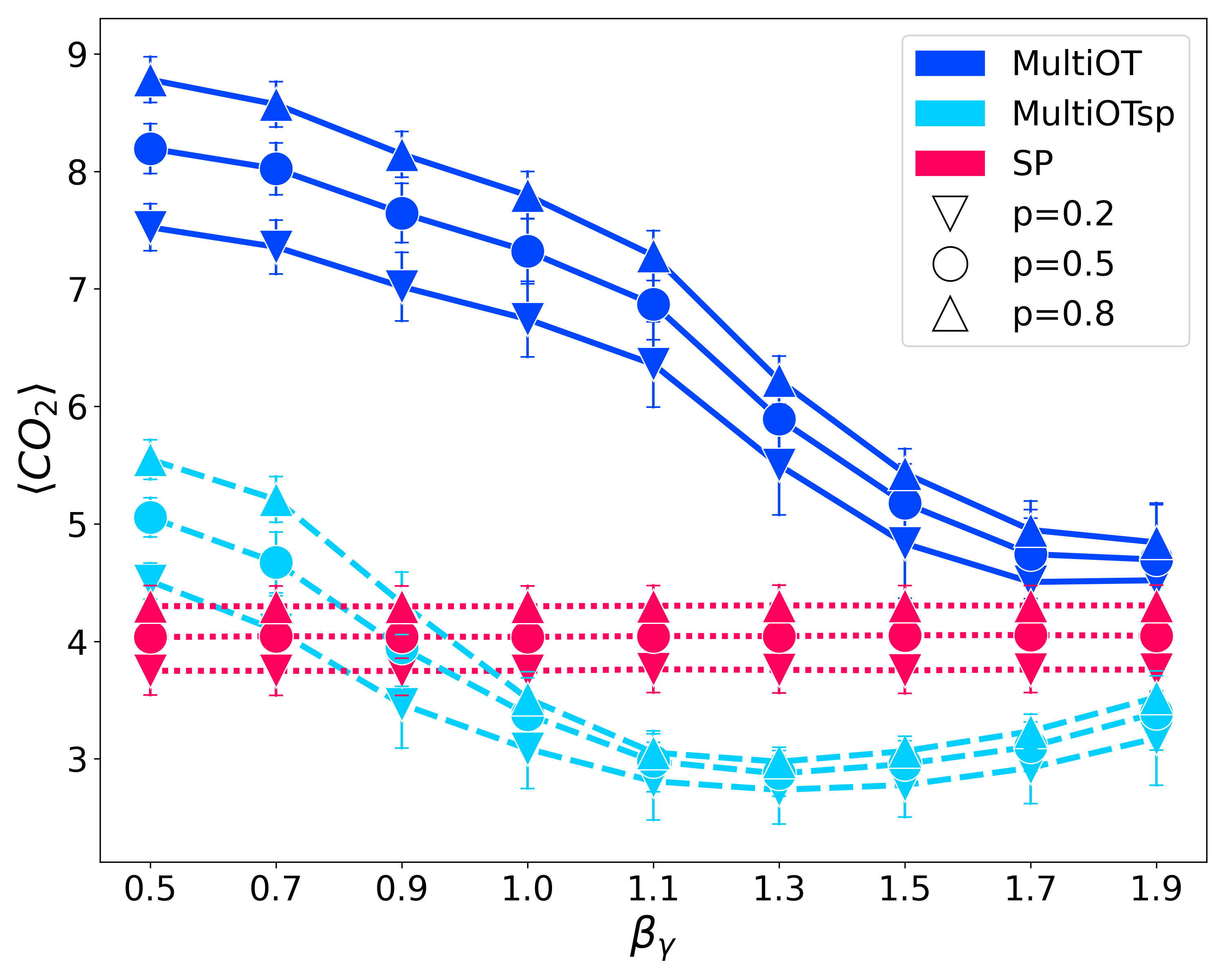}}
	\subfigure[]{\includegraphics[width=.47\linewidth]{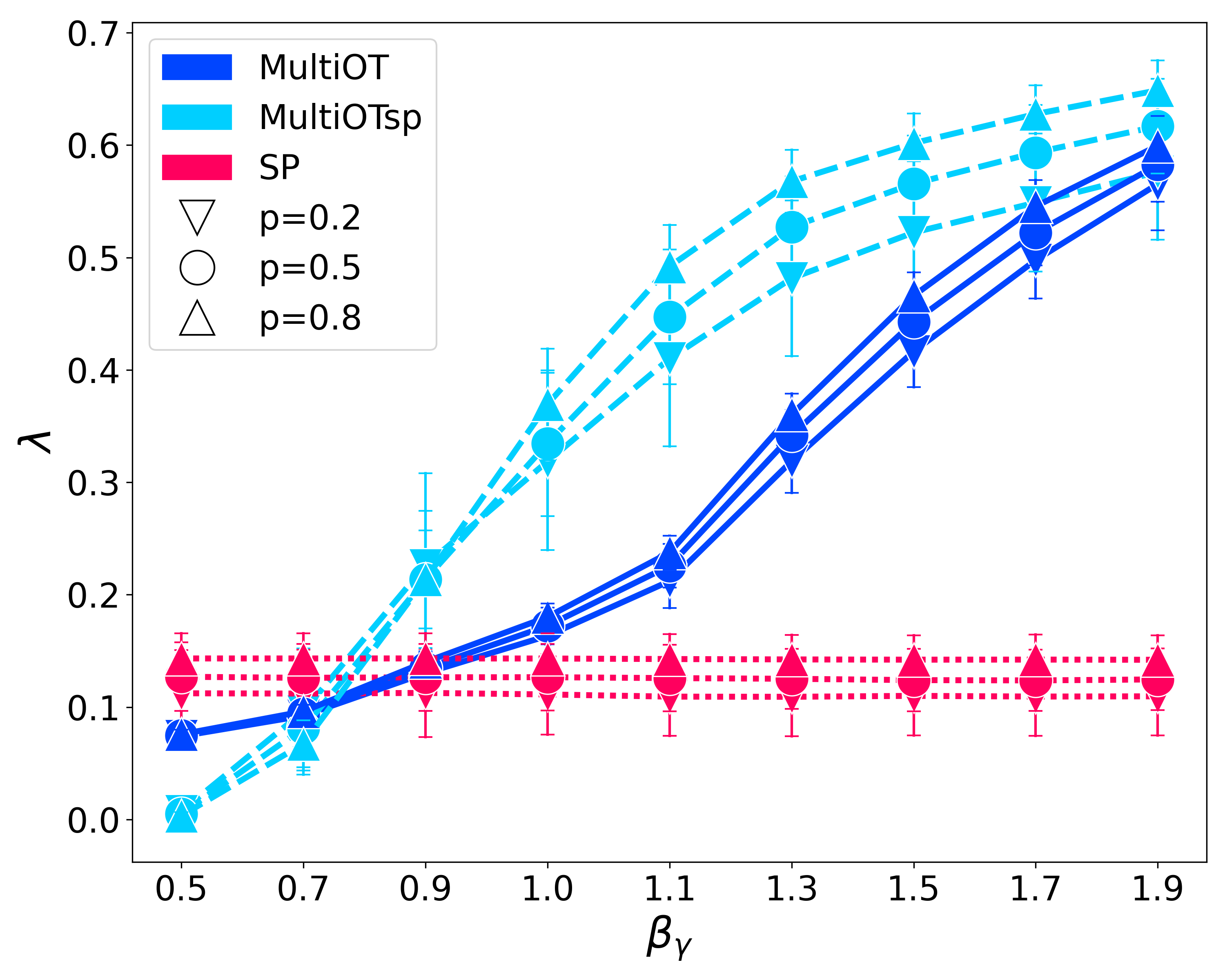}}
	\subfigure[]{\includegraphics[width=.47\linewidth]{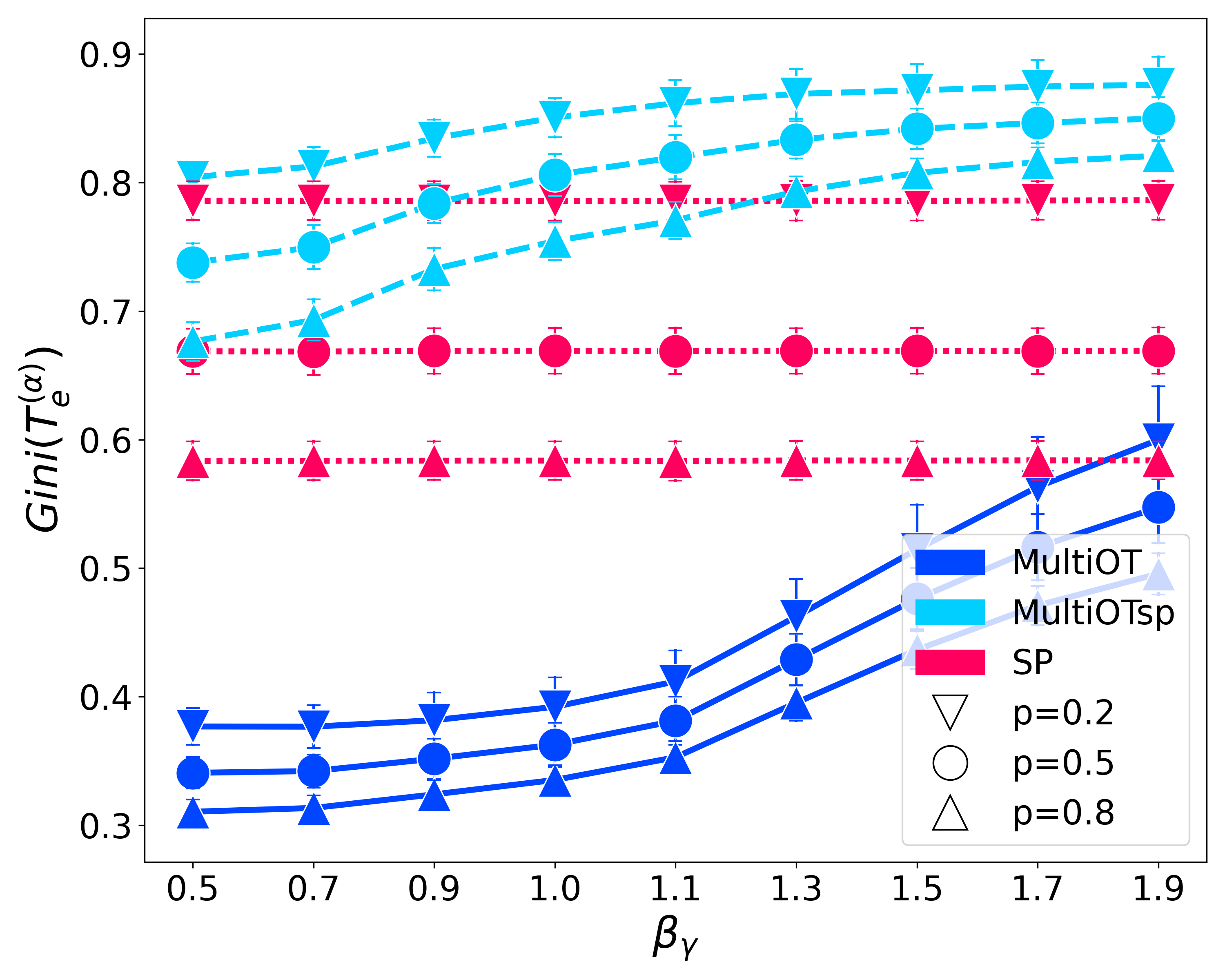}}
	\caption{Additional results on synthetic data for varying $p$. The results are averages and standard deviations over $20$ different network realizations with $100$ independent samples of origin-destination pairs on each network realization. Other parameters used here are $\beta_{\alpha}=0.5$, $N_{\alpha}=300$, $N_{\gamma}=60$.}
	\label{fig:apxExtrap}
\end{figure}

We set $p=\{0.2, 0.5, 0.8\}$ to capture different traffic demand scenarios, where $p=0.2$ and $p=0.8$ correspond to having the majority and minority of the passengers with a monocentric destination.  We show in \Cref{fig:apxExtrap} the performance of the algorithms in terms of the same metrics investigated in the main manuscript.   All displayed results have the same settings described in \Cref{sec:empirical}.

\bibliographystyle{apsrev4-2}
\bibliography{bibliography}

\end{document}